# Improving production process performance thanks to neuronal analysis

**M. Noyel\*,\*\*, Philippe Thomas\*\*,
P. Charpentier\*\*, A. Thomas\*\*, B. Beauprêtre\***

*\*ACTA Mobilier Parc d'activité Macherin Auxerre Nord 89470 MONETEAU – France (e-mail: mnoyel@acta-mobilier.fr and bbeauprêtre@acta-mobilier.fr)*

*\*\*Centre de Recherche en Automatique de Nancy (CRAN-UMR 7039), Lorraine-Université, CNRS Faculté des sciences et techniques, BP 70239 54506 VANDOEUVRE-LES-NANCY Cedex – France (e-mail : philippe.thomas@univ-lorraine.fr, patrick.chaprentier@univ-lorraine.fr and andre.thomas@univ-lorraine.fr).*

**Abstract:** Product quality level is become a key factor for companies' competitiveness. A lot of time and money are required to ensure and guaranty it. Besides, motivated by the need of traceability, collecting production data is now commonplace in most companies. Our paper aims to show that we can ensure the required quality thanks to an "on-line quality approch" and proposes a neural network based process to determine the optimal setting for production machines. We will illustrate this with the Acta-Mobilier case, which is a high quality lacquerer company.

## 1. INTRODUCTION

In the mass customization context, the main cause of the lack of control in the manufacturing processes is linked to the quality problem. Policies such as Total Quality Management (TQM) are defined to control it in the most efficient way. The American Production and Inventory Control Society defines Total Quality Management as "A management approach to long-term success through customer satisfaction. TQM is based on the participation of all members of an organization in improving processes, goods, services, and the culture in which they work". This definition is actually not far from the definition of Just in Time (JiT) which the same institution presents as "A philosophy of manufacturing based on planned elimination of all wastes and on continuous improvement of productivity. One of the main elements of JiT is to improve quality to zero defects. In the broad sense, it applies to all forms of manufacturing". JiT and TQM are also 2 major concepts related to Lean manufacturing –LM- (Vollmann *et al*., 84). Quality needs to be source-controlled. It means that we must verify quality level as close to the defect as possible. Taguchi is the first to consider the quality before the product origination with the set-up parameters control. But, this approach present the drawback to be off-line. Production processes control requires to set up a reactive control assistance system as well as quality control. To do that, the Deming wheel (Plan-Do-Check-Act) describes the different steps of the continuous improvement process where a fifth step, observation, may be added (OPDCA cycle) in order to highlight the need of measurement and analysis first. Different tools may be used to improve quality such as the 7 basic quality tools (Ishikawa chart, check sheet, control charts, histogram, Pareto chart, scatter diagram, stratification), or experimental designs. These tools, along with the Taguchi Method, are well known in the industry but they all suffer from having to be performed, "off line". We propose here an on-line method to improve production processes performance using neural networks to extract knowledge coming from production data and consequently, to understand production parameters effects, in line with the Taguchi Method. The main benefits of our approach are to exploit data collected in real conditions without the need of specific experiments, and to allow an "on-line" quality control. First, we will see globally the processes control and then we will discuss the proposed methodology to control quality processes. We will illustrate this methodology with an application case in the company Acta-Mobilier. At last, conclusion and outlook are presented.

## 2. PRODUCTION PROCESSES CONTROL

Processes control and continuous improvement are essential points for ISO 9001. Quality process is often the first to be studied because it directly impacts productivity. It is also the predominant concept in LM. Indeed, Lyonnet (2010) has summarized the different approaches of LM to identify points of action in which quality plays a major role.

### 2.1 Quality versus Productivity

Productivity and quality improvements are inseparable because quality has consequences on production and control. Overall, it is possible to classify non-quality consequences on, at least, the following points:

- Either it leads to a complication of the physical flow of products that must be repaired.
- Either it causes production and delivery delays. This is the case of products which need to be made again completely and which will not be ready for shipment.

To summarize, products flow is disturbed by products that must be reworked and those that should be treated urgently. This anomalous flow pattern fluctuates with non-quality rate. These are the fluctuating and unpredictable disturbances that complicate the production management and undermine the development of product control system. Monitoring and maintaining the quality level is a priority before one can be interested in production control.

*2.2 How to control quality?*

In order to control quality, we must first understand it. That is why it is necessary to know precisely the factors affecting quality. The manufactured products quality relies on the effects of many factors which can be classified in different ways. For example, we can classify them according to 6M (Ishikawa, 1986) of the Ishikawa chart (Machine (technology), Method (process), Material, Man Power, Measurement (inspection), Milieu (environment). When considering their controllability, we can retain 4 of the 6M:

- Environmental factors (Environment-Milieu) such as temperature or humidity. These factors are generally low- or non-controllable because, although they are easy to measure, it may be difficult to enslave them.
- Technical factors (Machine and Method) resulting primarily from the machine state during operations. These factors are controllable because they correspond exactly to the machine settings.
- Human factors (Man Power) in manual operations. They are difficult to take into account because they sometimes vary consequently between operators. So far,the attempts to control human factors (establishing standards, poka yoke) have limitations and constraints.

When we correctly analyze the influential factors, we can focus on the challenge for quality control. Generally, the "zero defect" can be obtained in 2 ways:

- By optimizing the initial settings of various factors.
- By drifts monitoring and prevention.

Drifts monitoring and prevention lead for example to the implementation of standards to oversee human factors, or preventive maintenance plans in order to limit technical factors drifts. Different ways may be used in order to optimize the initial settings of technical factors. Historically, there are 3 approaches to manage quality. The first one, and also the easiest, is to use the 7 basic quality tools (Ishikawa chart, check sheet, control charts, histogram, Pareto chart, scatter diagram, stratification). In this approach, the finished parts are a posteriori controlled and improvement propositions are performed by using expert knowledge. In a second approach, the main goal is to control the process and no longer the finished parts in order to tune the technical factors off-line by using experimental design methods. Vivier (2002) shows that it is possible to make real experimental designs as well as virtual ones, primarily by simulation. The Taguchi method, with experimental designs especially studied for industrial applications, allows great savings in time and money by significantly reducing the needed number of experiments to test the different influent factors and by organizing and sequencing experiments in order to optimize settings time thanks to a fractional plan. However, even with this low-costs method, it happens the decision makers refuse to launch experiments, mainly because of 3 things:

- The need for time. It is sometimes impossible to secure time to make experiments. For example, on a bottleneck machine where the charge rate is optimized, immobilizing the machine means a loss of production with a direct impact on the delivery rate.
- The need for material. Experiments on a machine consume semi-finished products. The cost of this semi-finished products has to be taken into account for calculating the experiment's cost. Besides, experiments cannot be done with products destined for customers because they usually produce defects.
- The obtained set up of technical factors may be optimal at the end of an experiment plans but it may not be robust to the evolution of the production process (change or modification of machines for example), because this approach is off-line by nature.

These 3 points often result in the giving up of true experimental designs to the benefit of small "test changes" whose real impact is often appreciated more than measured. Moreover, the Taguchi method has another significant drawback, even if you decide to go on with experiments, which is the difficulty of finding the optimum setting. Indeed, in cases where there are to much factors, where the amount of non-controllable factors is too large, or where there are too many interactions, could make it difficult to determine the optimal level for each factor. For example, if temperature is an important factor, it may be difficult to take into account its full range of variation and to preserve optimal control in the same time. The third and last approach to control quality is the on-line method. Perhaps, on-line control are often set aside because of the lack of resources or expertise. Indeed, the crucial point for these methods is to produce a model of the system as close to reality as possible in order to have it behave quite the same way in all possible situations. In this paper, we propose an on–line control method to ensure quality control thanks to a data mining process.

## 3. METHODOLOGY FOR INDENTIFY AND MASTER PROCESSES

Most companies work with a computerized production monitoring witch collects production information (time and material consumption) in a semi-automated or automated way. Knowledge Discovery in Data (KDD) process may be performed in order to identify valid, novel, useful and understandable patterns by exploiting the amount of data collected. A KDD process is performed in different steps (Patel and Panchal, 2012). We assign a letter to each step in order to refer more easily to them later.

- Selection: obtain data from various sources (a),
- Preprocessing: cleanse data (b),

- Transformation; convert to common format, transform to new format (c),
- Data mining: obtain desired results (d),
- Interpretation/Evaluation/Presentation: present results to user in meaningful manner (e).

The 2 main steps are selection (a) and data mining (d). Data mining which is the core part of KDD is the process of analysing data and summarize it into useful information. Different approaches can be used to perform it such as artificial intelligence, machine learning, statistics and database systems. Data mining may perform different tasks:

- Classification: maps data into predefined group or classes,
- Regression: maps data from an input space to an output space,
- Clustering: groups similar data together into clusters,
- Summarization: maps data into subsets with associated simple descriptions,
- Link analysis: uncovers relationships among data.

In a quality monitoring problem, the data mining must perform a classification of data into 2 classes: defect occurrence and no defect occurrence.

### 3.1 Selection (a), preprocessing (b) and transformation (c) of data

An important task in KDD process is the data collection (a). It is possible to collect the values of the different factors that influence quality in the same way. If data is correctly stored, we can analyze it in order to achieve a workstation experimental design and obtain the same results, it means detecting low- or non-influents factors and setting other factors at the best level. So, it seems to be possible, without additional cost, to prevent defect apparition and to specify ameliorative settings from the input data retained. The availability of data is a crucial point for the quality analysis.

Data collection is certainly the most restrictive point. Indeed, it implies that a brainstorming (identical to that achieved before the implementation of an experimental Taguchi plan in order to list the factors affecting quality) has been done and it must then wait several months (and ideally a year) to gather enough data to cover all cases of use of the machine. On the other hand, data collection is often seen as a waste of time because the operator must stop his work to write information not used directly in production. Especially since this collection should be done at 2 levels:

- Upstream of the workstation to write the manufacture's condition. It is at this point that we must collect the factor levels that are considered influential.
- Downstream of the workstation for defects input.

That is why it is very important to work on collecting data with the aim of making interfaces as intuitive and requiring as little time input as possible. Downstream of the workstation, one will collect the percentage of occurrences of each type of defect existing, focusing on computer unit counting and automatic selection of outgoing lot. Upstream of the workstation, it will collect the value of influencing factors, focusing on automatic data input as often as possible to save time for the operator and gather more reliable information. For companies that have already established production monitoring, interfaces for data input/output (such as the time taken to complete a lot, the amount of material consumed, the program used...) have already been set up. It is necessary to consolidate the 2 types of input (production information and quality factor values) to ease the work of operators. Fortunately, due to traceability needs, companies often collect lots of data which can be exploited during the KDD process. This data must be preprocessed (b) in order, for example, to synchronize the different database, delete evident outliers, and digitize qualitative data as color (c)...

### 3.2 Datamining (d)

As shown by Agard and Kusiak (2005), the volume of data to be analyzed is often weighty. Companies are used to archive, primarily for traceability reasons but they rarely use their well of information and only as indicators for real-time management methods. Management and quality improvement by data mining methods has been discussed in (Kusiak, 2001). Data mining is the part of Knowledge Discovery in Data which consists in analyzing data in order to summarize it into useful information. In our case, data mining should perform a classification of data into 2 classes: defect occurrence and no defect occurrence. To do that, different tools may be used such as Naïve Bayes, Decision tree, Support Vector Machine (SVM), neural networks (NN)... Decision tree is faster to classify data but does not work well with noisy data (Patel and Panchal, 2012). So in the case of industrial data, the use of this approach is not pertinent. Naïve Bayes is dedicated to the treatment of discrete data and the use of continuous need to perform a discretization of these data. Support Vector Machine and Neural network both use very close concepts which lead to very close results. Sometimes, SVM gives better results (Meyer et al., 2003), sometimes it is NN (Paliwal and Kumar, 2009; Hajek and Olej, 2010). In our case, in order to build the system's model, analysis is done directly through a neural network. The collected data is divided into 2 parts. One will be used for learning, and so for creating the model, while the other will be used for validation. Which means checking that our model has the same behaviour as the physical system. The structure of the neural network used here is recalled in the next section.

## 4. STUDY CASE

### 4.1 Presentation of the company and its processes

The company Acta-Mobilier follows perfectly the context of maintaining quality issues. It produces high quality lacquered panels made in MDF (Medium Density Fiberboard) for kitchens, bathrooms, offices, stands, shops, hotel furniture... In its quest for continuous improvement, the company is certified ISO 9001, ISO 14001, OHSAS 18001 and implements the Kaisen process (5S). We recall that one of the

4 points that summarize the Kaisen ideology refers only to quality with the goal of eliminating defects completely.

The activity is divided into 5 workshops. Each workshop is likely to generate defects and we should include a quality control at all stages of the manufacturing process to be able to control the rate of defects. However, the main defects are generated in the lacquering step. So, this paper focuses on the robotic lacquering workstation. Even if this workstation is free of human factors, the production quality is unpredictable (we cannot know if there is a risk that products will have defects) and fluctuates (the percentage of defects may be of 45% one day and down to 10% the next day without changing the settings).

*4.2 The neural network chosen: The multilayer perceptron*

We propose the use of neural networks (NN) to model systems on which we can realise the Taguchi experiments plan or global plans with an on-line control method in order to determine the best settings for each factor. The advantages of this approach are:

- Exploitation of real data without carrying out dedicated experiments as in experiments plan because we can use the database collected during the production.
- Simple implementation of the approach because the neural model design is partially automated.
- On-line tuning of the quality monitoring process by using actual production data in order to improve and adapt the process to change.

In our approach, the neural model is performed by using production data representatives of all the conditions encountered in the past and so, it can adapt itself to these changeable conditions. This model is able to provide lower and upper limits for each controllable factor settings based on all non-controllable factors. We are looking to implement a tool at the input of key machines, able to determine in real time these landmarks, and to verify that factors are well set within these landmarks.

The multilayer perception (MLP) seems to be the NN best suited to our case. Works of Cybenko (1989) and Funahashi (1989) have proved that a MLP with only one hidden layer using a sigmoïdal activation function and an output layer can approximate all non-linear functions with the wanted accuracy. Its structure is given by:

$$z = g_2\left(\sum_{i=1}^{n_1} w_i^2 \cdot g_1\left(\sum_{h=1}^{n_0} w_{ih}^1 \cdot x_h^0 + b_i^1\right) + b\right) \quad (1)$$

where $x_h^0$ are the $n_0$ inputs of the NN, $w_{ih}^1$ are the weights connecting the input layer to the hidden layer, $b_i^1$ are the biases of the hidden neurons, $g_1(.)$ is the activation function of the hidden neurons (here, the hyperbolic tangent), $w_i^2$ are the weights connecting the hidden neurons to the output one, $b$ is the bias of the output neuron $g_2(.)$ is the activation function of the output neuron and $z$ is the network output. Because of this, the problem is to obtain a probability of defect occurrence, $g_2(.)$ being chosen sigmoïdal.

Now, only the number of hidden neurons is always unknown. In order to determine it, the learning starts from an overparametrized structure. A weight elimination method is used to remove spurious parameters (Setiono and Leow, 2000). The learning of the MLP is performed in 3 steps:

- Initialisation of weights (Nguyen and Widrow, 1990).
- Learning of parameters by using Levenberg-Marquard algorithm with robust criterion (Thomas *et al.*, 1999).
- Weights elimination (Setiono and Leow, 2000).

*4.3 Results*

Following the brainstorming about the factors influencing the quality level, we were able to classify them in 3 categories specified in paragraph 2.2. We could imagine the potential drifts and associate a type of defect. This preliminary work results of expert knowledge. Only after the complete study can we know accurately which factors affect which defects. To illustrate these categories, we present in Table 1 some trivial examples corresponding to experts ideas.

**Table 1. Examples of factors, deviations and associated defects for 3 types**

| Factors types | Control-lability | Factor | Deviation | Defect |
|---|---|---|---|---|
| Environmental | Middle | Temperature | Increase at midday → drying time too short | "Microbubbling" |
| Technique | Very good | Dust suction power | Decrease due to compressor fatigue → rest of dust before lacquering | Grains |
| Human | Low | Spray speed | Slowdown due to a pain → dry lacquer covering up | Bands phenomenon |

Thanks to a production and quality management system, data corresponding to the factors studied since February 2012 are collected. Upstream of the robotic lacquering workstation, experts decided to collect factors such as load factor, number of passes, time per table (lacquering batches), liter per table, basis weight, number of layers, number of products and drying time.

We could add to these technical factors environmental ones such as temperature, atmospheric pressure and humidity. According to the experimental design approach, we can classify these variables into 2 types of factors: internal ones (load factor, number of passes, time per table, liter per table, basis weight, number of layers, number of products and drying time) and external ones (temperature, humidity, pressure). 2 factors (passes number and number of layers) are discrete ones that can each take 3 states and that are binarized (Thomas and Thomas 2009).

We have 15 inputs to apply to the NN which 6 are binary ones. Downstream of the machine, we were able to detect up to 30 different types of defects. The first works with the NN prediction concerning a type of defects: "Stains on back."

We have a total of 2270 data we will split into 2 data sets for identification (1202 data) and validation (1068 data). First, learning is achieved by exploiting 25 neurons in the hidden layer. Pruning phase can then eliminate spurious inputs and hidden neurons. This is done 100 times with different initial weight sets to avoid problem of trapping in local optimum. After pruning, 6 hidden neurons and 1 input (passes number) are eliminated. This means that the passes number has very little influence on the defect concerned and so does not need to be taken into account. This "data processing" is used to simplify the model and avoid over fitting problem.

During the validation phase, we therefore compare the results of the NN with the real defects detection. For the defect "Stains on back", we know that it occurs 127 times on the 1068 data validation set. The NN can detect 112 defects which lead to a non-detection rate of 11.8% (Fig. 1). The proportion of false positive is 19.2%, which may be partly explained by the fact that some defects haven't been identified out of the machine.

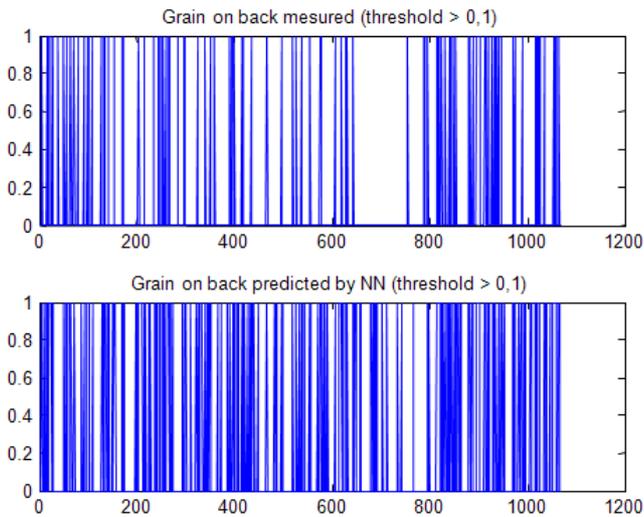

Fig. 1. Comparison of "Grain on back" defects detected at the robotic lacquering workstation output (top graph) and defects predicted by the NN (bottom graph)

It is obvious that this defect is largely explained by the archived operating conditions and it's possible to use NN upstream of the workstation to prevent the risk of "Stains on back". This NN is improved and adapted on line by using current production data and on line learning algorithm. We repeated the experiment with other defects. Obviously, it would first analyze 20% of defects that create 80% of non-quality (Pareto). However, there are defects which cannot be predicted with the NN in the conditions described above. This is for example the case for "knock" where you get 73% non-detection for 11% false positives. The non predictable defects depend certainly on other factors that we need to determine and collect if we want to predict them. In total, on the 30 identified defects, 7 can be partially explained using the variables collected. For the predictable defects, there are 2 possible approaches:

- Warning. By analyzing the inputs through the NN, it becomes possible to predict defects occurrence and report it when conditions are met to create risk.
- Limitation. By using NN to limit the input factors by upper and lower limits and prohibit the production lot when one of the inputs is outside these limits. If it's a controllable factor, operators can modify it to allow production. Otherwise, the production lot will be rejected. It must be scheduled later when conditions become acceptable. To do that, factors must be classified into controllable ones (load factor, basis weight, drying time, liters per table), non-controllable ones (temperature, humidity, pressure) and protocols (number of passes, time per table, number of layers, and number of products). The NN is then used instead of the real system to perform experiments that achieve an entire plan without increasing the cost. The results, however, still to be validated on the real system.

We present the results of an entire plan in which 10 levels were chosen for the 3 controllable factors (Fig. 2). For protocol factors, we set the number of passes and layer to 1; the time per table, and the liter of lack to their average values; and the number of products to its median value. We also fixed non-controllable factors values to their average values.

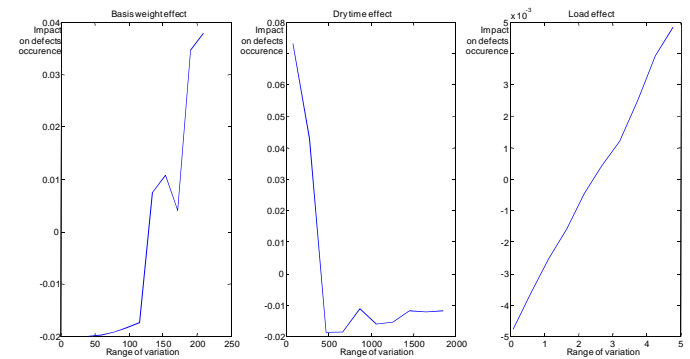

Fig. 2. Experience plan results by using NN model.

These figures show that the load factor has a relatively small impact on the defects occurrence. However, the increase in weight has a significant effect. High basis weights tend to create more defects. Drying time has also an impact because too short drying time greatly increases the occurrence of defects. In order to obtain the same results by using an experiments plan, we need to use 5 modalities for drying time and basis weight and 2 for the load factor that lead to many experiments even using Taguchi plan. These preliminary results need to be confirmed by taking into account the variation of non-controllable factors (temperature, pressure and humidity) and validating the results on the real system.

*4.4 Direct advantages : defect rate decrease*

In our study, the most recurrent defect concerns the "grains on face". Unfortunately, we have seen that this defect was not predictable with our inputs. However, in occurrence order, the second default is the "microbubbling" which still represents 12% of all defects with a average occurrence of 72 defective products per week, which means 1.1% of the production. As this defect is difficult to recover and it usually needs to sand the product again and repeat the lacquering. Avoiding this defect by eliminating the need of operators as soon as the conditions are met or by indicating the change of setting would make 20000 € saving per year.

*4.5 Indirect advantages: reduction in the number of experiences*

The saving shown in the previous paragraph is a net saving. Indeed, it required no experience, so no material or time consumption at the concerned production workstation. To calculate the avoided costs, we still made the study by Taguchi experimental plan. Depending on the number of input factors and the number of modality (often 3) we had to move towards a Taguchi matrix with 36 experiments. To minimize costs, we may decide that each experience consists in lacquering a table where there are at least 3 products (semi-finished). In these conditions, the Taguchi plan will require the lacquering of 108 products on 36 tables. The cost of the Taguchi plan may be estimated by taking into account, time consumption, setting time, hourly cost for the robotic lacquering workstation, material consumption (semi-finished products), lacquer consumption... We can also add other costs that will result from the implementation of the Taguchi plan, for example time for teams briefing, time to summarize data, the cost of moving away products from robotic lacquering to manual lacquering where the hourly cost is higher... with all this, we can estimate the overall cost of the Taguchi experiment plan to about 5000 €. We have a return on investment of a quarter with the Taguchi experiment plan versus a direct return on investment for the NN method.

## 5. CONLUSION AND OUTLOOK

It's important not to separate quality control and production management, first because the quality level has high impact on production management, and then because one of the key elements, we mean the factor values collected during the operation can be directly related to the production information collected (time and material consumption). Regarding the data processing, we have highlighted the benefits of using a neural network rather than a traditional Taguchi plan that still has the inconvenience of stopping the machine and inevitably consumes raw materials or semi-finished products. NN can lead to the results of experimental designs by getting rid of these problem areas and so generate net savings without subtracting the costs of experiments. Their implementation in the workshop will enable more reliable product flow and allow the works on production control. These data mining tools have their place at the heart of the competitiveness business struggle.